\documentclass[11pt,a4paper]{article}
\usepackage{amssymb,amsmath,palatino,amsfonts,xcolor}
\usepackage{graphicx,subfigure}
\newcommand\be{\begin{equation}}
\newcommand\ee{\end{equation}}
\newcommand\bea{\begin{eqnarray}}
\newcommand\eea{\end{eqnarray}}
\newcommand\ket[1]{|#1\rangle}

\newcommand{\kb}[2]{{|#1\rangle} \! {\langle #2|}}
\newcommand{\bk}[2]{{\langle #1|}{#2\rangle}}

\title{\large \bf Maximal Sets of Mutually Unbiased Quantum States in Dimension Six 
}
\author{\normalsize Stephen Brierley and Stefan Weigert \\
        \normalsize      Department of Mathematics, University of York\\
        \normalsize      Heslington, UK-York YO10 5DD\\
        \normalsize \tt{sb572@york.ac.uk,slow500@york.ac.uk}}
\date{\normalsize August 2008}

\begin{document}

\maketitle

\begin{abstract}
We study sets of pure states in a Hilbert space of dimension $d$ which are \emph{mutually unbiased} (MU), that is, the moduli of their scalar products are equal to zero, one, or $1/\sqrt{d}$. These sets will be called a MU \emph{constellation}, and if four MU bases were to exist for $d=6$, they would give rise to 35 different MU constellations. Using a numerical minimisation procedure, we are able to identify only 18 of them in spite of extensive searches. The missing MU constellations provide the strongest numerical evidence so far that no seven MU bases exist in dimension six.     
\end{abstract}


\section{Introduction}

The \emph{dynamics} of an autonomous Hamiltonian system with a single degree of freedom differs considerably from that of a system with two or more degrees of freedom. Nontrivial interactions among the degrees of freedom usually lead to an effectively unpredictable time evolution. From a \emph{kinematical} point of view, however, there is not much of a difference: the composite system simply inherits the fundamental symplectic structure of its constituents.

Schwinger associates one degree of freedom with a quantum system whenever the dimension $d$ of its Hilbert space is a prime number \cite{schwinger60}. Quantum systems with two or more degrees of freedom are obtained by tensoring copies of these building blocks. Our classically trained intuition wants to make us believe that the kinematics of composite quantum systems will not depend on the dimensions of the building blocks. In other words, we expect that composite quantum systems with dimensions $d_1 =2 \times 3$ and $d_2 =3 \times 3$, for example, are structurally identical. 

The concept of \emph{mutually unbiased} (MU) bases \cite{ivanovic81} appears to invalidate this expectation since complete sets of MU bases seem to exist in prime power dimensions only. They are an important, physically motivated tool allowing one to reconstruct quantum states with optimal efficiency \cite{wootters+89}. Given a quantum system of dimension $d$, a \emph{complete set of MU bases} in $\mathbb{C}^d$ consists of $d(d+1)$ pure states $\ket{\psi^{b}_{j}}$, $b=0,1,\ldots,d$, $j=1, \ldots,d$, which satisfy the conditions
\begin{equation}  \label{MUB conditions}
\left| \bk{\psi_{j}^{b}}{\psi_{j^\prime}^{b^\prime}} \right| 
 = \left\{ 
\begin{array}{ll}
\delta _{b b^\prime} & \quad \mbox{if $b = b^\prime$} \, , \\ 
\frac{1}{\sqrt{d}} & \quad \mbox{if $b \neq b^\prime $} \, .  
\end{array}
\right.
\end{equation}
Thus, the states form $(d+1)$ orthonormal bases, and scalar products between states taken from different bases have constant modulus. If the dimension $d$ is a prime or the power of a prime, complete sets of MU bases do exist, and it is impossible to have more than $(d+1)$ such bases. For \emph{composite} dimensions $d=6,10,12, \ldots$, however, their existence poses an open problem despite many efforts reviewed in \cite{planat+06}. 

The purpose of this paper is to systematically search for \emph{subsets} of complete sets of MU bases which we will call \emph{MU constellations}. Essentially, a MU constellation consists of groups of $d$ or fewer vectors having scalar products as in (\ref{MUB conditions}). Three MU bases, known to exist in any dimension $d$, are a well-known example of a MU constellation. It has been conjectured \cite{butterley+07} that four MU bases, another MU constellation, do not exist in  dimension six. The non-existence of a MU constellation consisting of three MU bases plus one additional vector, related to the Heisenberg-Weyl group, has been shown in \cite{grassl04}. There are, however, many other entirely unexplored MU constellations.   

We focus on MU constellations in dimension six, the smallest value for $d$ not equal to the power of a prime. We will find that many MU constellations with less than $42$ states are highly unlikely to exist. These missing MU constellations will provide the strongest numerical evidence so far that no seven MU bases exist in dimension six. Based on our findings, we will formulate a simple argument to explain the observed lack of MU constellations beyond three MU bases. 

This paper is organised as follows. In the next section, we introduce the concept of  MU constellations and embed them in well-defined searchable spaces. Then, in Section \ref{numerical search} the search for MU constellations is cast into the form of a numerical minimisation. Section \ref{Results} describes the results of the searches, and they will be discussed in the final section.

\section{Constellations of quantum states in $\mathbb{C}^d$ \label{Partial MUBs}}

In this section we define \emph{mutually unbiased constellations} of quantum states and we embed them in appropriate spaces to search for them. 

\subsection{Mutually unbiased constellations}

A \emph{MU constellation} in $\mathbb{C}^d$ consists of $(d+1)$ sets of $x_b$ pure states $\ket{\psi^{b}_{j}}$, $b=0,1,\ldots,d$, $j=1, \ldots,x_b$, which satisfy the conditions (\ref{MUB conditions}). The $(d+1)$ integers $x_b \in \{ 0, \ldots , d-1\}$ specify all possible types of MU constellations which will be denoted by 
\be \label{MUconstellationnotation}
\{ x \}_d \equiv  \{x_{0},x_{1},\ldots ,x_{d}\}_{d} \, , 
\quad
x \in \left( \mathbb{Z} \mbox{ mod } (d-1)\right)^{d+1} \, .
\ee
If the number $x_b$ in a MU constellation $\{ x \}_d$ equals zero, it corresponds to an empty set and will be suppressed. For example, $\{2,1,2,0\}_{4} \equiv \{2,1,2\}_{4}$ denotes a MU constellation in $\mathbb{C}^4$ which consists of two pairs of orthonormal vectors and one single vector. Since the ordering of the bases within a constellation will be irrelevant, we arrange them in decreasing order, using the shorthand $x^{a}$ if there are $a$ bases with $x$ elements: $\{2,1,2\}_{4}$ thus becomes $\{2^{2},1\}_{4}$. 

The numbers $x_b$ are limited to $(d-1)$ since there is only one way to complete $(d-1)$ orthonormal vectors to a basis of the space $\mathbb{C}^d$, apart from an irrelevant phase factor. More explicitly, the complement of $(d-1)$ orthonormal vectors $\ket{\psi_{j}} \in \mathbb{C}^d, j = 1, \ldots , d-1$, is a unique one-dimensional subspace spanned by $\ket{\psi_\perp}$, say. Due to the completeness relation for an orthonormal basis, the projector on this subspace must have the form
\be \label{uniqueprojector}
\kb{\psi_\perp}{\psi_\perp} 
=\mathbb{I}_d -\sum_{j=1}^{d-1} \kb{\psi _{j}}{ \psi _{j}} \, ,
\ee
where $\mathbb{I}_d$ is the identity in $\mathbb{C}^d$. 

The completion of $(d-1)$ orthonormal vectors into a basis is \emph{consistent} with the conditions of mutual unbiasedness (\ref{MUB conditions}). The identity (\ref{uniqueprojector}) implies that the state $\ket{\psi_\perp}$ is MU with respect to any vector $\ket{v}$ satisfying $|\bk{\psi_j}{v}|=1/\sqrt{d}$, hence any MU constellation containing the states $\{ \ket{\psi_{j}}\}$ remains MU if the state $\ket{\psi_\perp}$ is added to the set.

MU constellations in $\mathbb{C}^d$ are a \emph{partially ordered set} with respect to the relation $\leq$ defined by 
\be \label{partialorder}
\{ x \}_{d}\leq \{y \}_{d} \Longleftrightarrow 
x_{b}\leq y_{b} \, , \quad \mbox{ for all } b=0,1,\ldots , d \, . 
\ee
The ordering refers only to the \emph{number} of vectors in each basis; it does not imply any relation between the subspaces spanned by the vectors in corresponding 'partial bases' of the constellations $\{x \}_{d}$ and $\{ y \}_{d}$. If (\ref{partialorder}) holds, we will say that $\{ y \}_{d}$ \emph{contains} $\{x \}_{d}$; alternatively, $\{x \}_{d}$ is said to be \emph{smaller} than $\{ y \}_{d}$. For example, the MU constellation $\{2^{2},1\}_{4}$ is contained in four MU bases $\{3^4\}_4$ because  
\be \label{d=3example}
\{2^{2},1\}_{4} \leq \{3^4\}_4
\ee
is true. The ordering induced by (\ref{partialorder}) is only partial since constellations such as $\{3,1\}_4$ and $\{2^2\}_4$ cannot be compared to each other. Thus, MU constellations possess a \emph{lattice structure} with a unique minimal element, $\emptyset$, and $(d+1)$ MU bases $\{ (d-1)^{d+1} \}_d$, if existing, provide a unique maximal element.  

Here is an important consequence of the lattice structure. A set of $k \in \{2,\ldots , d+1\}$ complete MU bases $\{(d-1)^{k}\}_{d}$ in dimension $d$ exists only if all \emph{smaller} MU constellations $\{ x \}_{d}$ exist, i.e. those with 
\be 
\{x_{0},x_{1},\ldots ,x_{k-1}\}_{d}\leq \{(d-1)^k \}_{d} \, , \quad  
0 \leq x_b \leq d-1\, , \quad b=0, 1, \ldots, k-1 \, .
\ee
Hence, if any MU constellation $\{x_{0},x_{1},\ldots ,x_{k}\}_{d}$ is found missing then $k$ complete MU bases cannot  exist. This observation has been exploited in \cite{butterley+07} where the unsuccessful numerical search for four MU bases, i.e. the MU constellations $\{5^4\}_6$, is used to argue that no seven MU bases exist for $d=6$. Similarly, it has been shown in \cite{grassl04} that it is impossible in $\mathbb{C}^6$ to extend two MU bases $\{5^2\}_6$ \emph{of a specific type} to the MU constellation $\{5^3,1\}_6$, excluding thus the existence of seven MU bases based on a specific construction. 

Evidence for the non-existence of any small MU constellation is evidence for the non-existence of the corresponding complete set of MU bases. This observation is crucial for the main thrust of this paper.

\subsection{Constellation spaces\label{constellationspace}}

In Section \ref{Results}, we will numerically search for all MU constellations $\{ x \}_6$ in $\mathbb{C}^6$ contained in $\{ 5^4\}_6$, i.e. in \emph{four} MU bases. To do this, we need to search through a space which is guaranteed to contain a specific MU constellation if it exists; at the same time, the search space should be as small as possible to maximize computational efficiency. From now on, we will only consider MU constellations which contain at least one complete basis, 
\be \label{restriction}
\{ x \}_d \equiv \{ d-1, x_1, \ldots , x_d \}_d \, , 
\ee
which is a mild restriction allowing that allows considerable simplifications. 

To associate an appropriate space with a given MU constellation $\{ x \}_d$ of type (\ref{restriction}), we will need to write it in \emph{dephased} form. Once dephased, its first $(d-1)$ vectors are given by those of the standard basis ${\cal B}_z$, while the components of the first vector of the second basis and the first component of each remaining vector are equal to $1/\sqrt{d}$. For example, upon dephasing a MU constellation $\{ 2^3,1 \}_3$, it takes the form 
\begin{eqnarray} \label{exampled=3,2221}
\left\{ \left( 
\begin{array}{cc}
1 & 0 \\ 
0 & 1 \\ 
0 & 0 %
\end{array}%
\right) \, , \frac{1}{\sqrt{3}}\left( 
\begin{array}{cc}
1 & 1 \\ 
1 & e^{i\alpha_{11}} \\ 
1 & e^{i\alpha_{21}} %
\end{array}%
\right) \, , 
\frac{1}{\sqrt{3}} 
\left( 
\begin{array}{ccc}
1 & 1 \\ 
e^{i\beta_{11}} & e^{i\beta_{12}} \\ 
e^{i\beta_{21}} & e^{i\beta_{22}} %
\end{array}%
\right) \, , 
\frac{1}{\sqrt{3}}\left( 
\begin{array}{c}
1 \\ 
e^{i\gamma_{11}} \\ 
e^{i\gamma_{21}} %
\end{array}%
\right) \right\} \, ,
\end{eqnarray} 
with specific values for the eight angles $\alpha_{11}, \ldots, \gamma_{21}$. It is shown in Appendix A that any given MU constellation of type (\ref{restriction}) can be written in dephased form by applying transformations which leave invariant the conditions (\ref{MUB conditions}).

Now it is straightforward to associate a \emph{space of constellations} with the MU constellation $\{ 2^3,1 \}_4$: the space ${\cal C}_4( 2^3,1)$ is defined as the set of vectors one obtains from (\ref{exampled=3,2221}) if the eight angles $\alpha_{11}, \ldots, \gamma_{21}$, are allowed to vary freely between 0 and $2\pi$. 
Each point in this space will be called a \emph{constellation} $[ 2^3,1 ]_4$, and it corresponds to a set of seven (not necessarily different) pure states in $\mathbb{C}^4$. Not all constellations $[ 2^3,1 ]_4$ are a MU constellation $\{ 2^3,1 \}_4$, but each MU constellation $\{ 2^3,1 \}_4$ is represented by at least one point of the space ${\cal C}_4(2^3,1)$.

In general, each MU constellation $\{ x \}_d$ is embedded in space ${\cal C}_d(x)$ of constellations $[x]_d$, defined in analogy to ${\cal C}_4( 2^3,1)$. Simply write down the dephased form of the MU constellation $\{ x \}_d$ at hand; then, varying the angles $\alpha_{11}, \dots,$ between 0 and $2\pi$, generates the space of constellations 
\be \label{defineconst}
{\cal C}_d(x) \ni [x]_d = [d-1,x_0, \ldots, x_d]_d \, .
\ee  
The space ${\cal C}_d(x)$ has the structure of a multi-dimensional \emph{torus} due to the periodicity of the angles used to parameterize it.
  
Let us now determine the dimension of the space ${\cal C}_d(x)$ associated with a MU constellation (\ref{restriction}). It contains  
\be \label{totalnumberofstates}
S = d-1 + s 
\ee
quantum states where 
\be \label{numberofvectorsinlastbases}
s=\sum_{b=1}^{d}x_{b} 
\ee
is the number of states in all groups but the first one. Since each of these vectors except the first one brings $(d-1)$ phases, the entire constellation $[ x ]_{d}$ depends on
\be \label{MUconstparameters}
p_d 
 \equiv p\left( [d-1,x_{1},x_{2},\ldots ,x_{d}]_{d} \right)
 = (d-1)(s-1)
\ee
\emph{independent real parameters}. For example, the constellation space ${\cal C}_d( (d-1)^{d+1})$ associated with $(d+1)$ complete MU bases has dimension $(d-1)(d^2-d-1)$. 

How many constraints does the requirement of mutual unbiasedness in (\ref{MUB conditions}) impose on the parameters of a constellation $[x]_d$? The states of a constellation are normalized, and the conditions on scalar products involving vectors of the first basis are satisfied by construction, so that there remains exactly one condition for each pair of different states taken from the last $d$ bases. Consequently, the number of \emph{constraints} is given by 
\be \label{numberofparametersgeneral}
c_d \equiv c_d \left( [d-1,x_{1},x_{2},\ldots ,x_{d}]_{d} \right) 
 = \frac{1}{2} s (s-1) \, .
\ee

The number of free parameters equals the number of constraints,
\be \label{equality}
p_d=c_d =(d-1)(2d-3) \, , 
\ee
whenever one considers a constellation with $s=2(d-1)$ states within the last $d$ groups. Constellations with $p_d=c_d$ will be called \emph{critical} ones, denoted by $[ {\bf x} ]_d$. Constellations of type $[d-1,x_{1},\ldots ,x_{{d}}]_{d}$ with more than $S=3(d-1)$ states are subjected to \emph{more} constraints than they possess free parameters. These \emph{overdetermined} constellations will be referred to as $[ \underline{x} ]_d$. 

\section{Numerical search for MU constellations\label{numerical search}}

This section explains the numerical method we use to identify MU constellations. The basic idea is to define a continuous function on the space of constellations ${\cal C}$ that takes the value zero if and only if the input is a MU constellation. We then search for the zeros of this function in the neighborhood of a large number of randomly chosen points in ${\cal C}$, using standard numerical methods.

\subsection{MU constellations as global minima} \label{MUglobal}

Suppose you want to find the MU constellation $\{ x \}_d$. To do so, consider the associated space of constellations ${\cal C}_d(x)$ which can be parameterized by $p_d$ angles denoted by $\vec{\alpha}= (\alpha_{1}, \ldots, \alpha_{p_d})^T$. Defining
\be \label{definitionofchi}
\chi _{jj^\prime}^{bb^\prime }
 =\left\{ \begin{array}{ll}
\delta _{jj^\prime} & \quad \mbox{if $b = b^\prime$}  \, ,\\ 
\frac{1}{\sqrt{d}} & \quad \mbox{if $b \neq b^\prime $} \, ,
\end{array}
\right. 
\ee
the non-negative function $F:\mathbb{R}^{p_d}\rightarrow \mathbb{R}$
\be \label{function}
F( \vec{\alpha} )
= \sum_{1\leq b \leq b^\prime}^{d}
  \sum_{j=1}^{x_b} 
  \sum_{j^\prime=1}^{x_{b^\prime}}
  \left( | \bk{\psi_{j}^{b}}{\psi_{j^\prime}^{b^\prime}} | 
                  - \chi_{jj^\prime}^{bb^\prime }\right) ^{2},
\ee
equals zero if and only if the input $[ x ]_d$ coincides with a MU constellation $\{x\}_d$.

It is thus possible, in principle, to prove the (non-) existence of a MU constellation by determining whether the smallest value of the function $F(\vec{\alpha})$ is non-zero. This means to identify its (possibly degenerate) \emph{global} minimum which, unfortunately, is not simple: the global minimisation of a nonlinear function such as a polynomial of fourth order in sufficiently many variables may already pose a NP-hard problem \cite{nesterov00}. A well-known strategy is to search for minima by starting from random initial points which, however, may turn out to be \emph{local} ones. By repeating the process sufficiently often, one will detect global minima as well---if they exist.

A numerical search along similar lines has been reported in \cite{butterley+07}, restricted, however, to the MU constellations $\{5^{4}\}_{6}$ and $\{5^{7}\}_{6}$, that is, four or seven MU bases. This limitation allows for a different parametrization which exploits the fact that complete bases in dimension $d$ are associated with $d$-dimensional unitary matrices. 

Note that the choice of the function $F(\vec{\alpha})$ is not unique.\footnote{We have also considered an everywhere differentiable variant of (\ref{function}) obtained by subtracting the square of $\chi _{jj^\prime}^{bb^\prime }$ from $| \bk{\psi _{j}^{b}}{\psi_{j^\prime}^{b^\prime}} |^{2}$. We noticed, however, that the success rate to find existing MU constellations is systematically lower.} The expression (\ref{function}) is convenient because efficient minimisation tools are available for a sum of squares. In particular, the Levenberg-Marquardt algorithm \cite{levenberg44,marquardt63}, often used in Regressional Analysis, cleverly switches between the method of steepest descent and the Gauss-Newton algorithm to speed up convergence. To search for zeros of the function $F( \vec{\alpha} )$, we use the function \texttt{optimize.leastsq} from the Open-Source Python package SciPy \cite{SciPy} which implements the LM-algorithm.

\begin{figure}[h]
\begin{center}
\includegraphics[viewport=260 25 700 350,width=8.5cm]{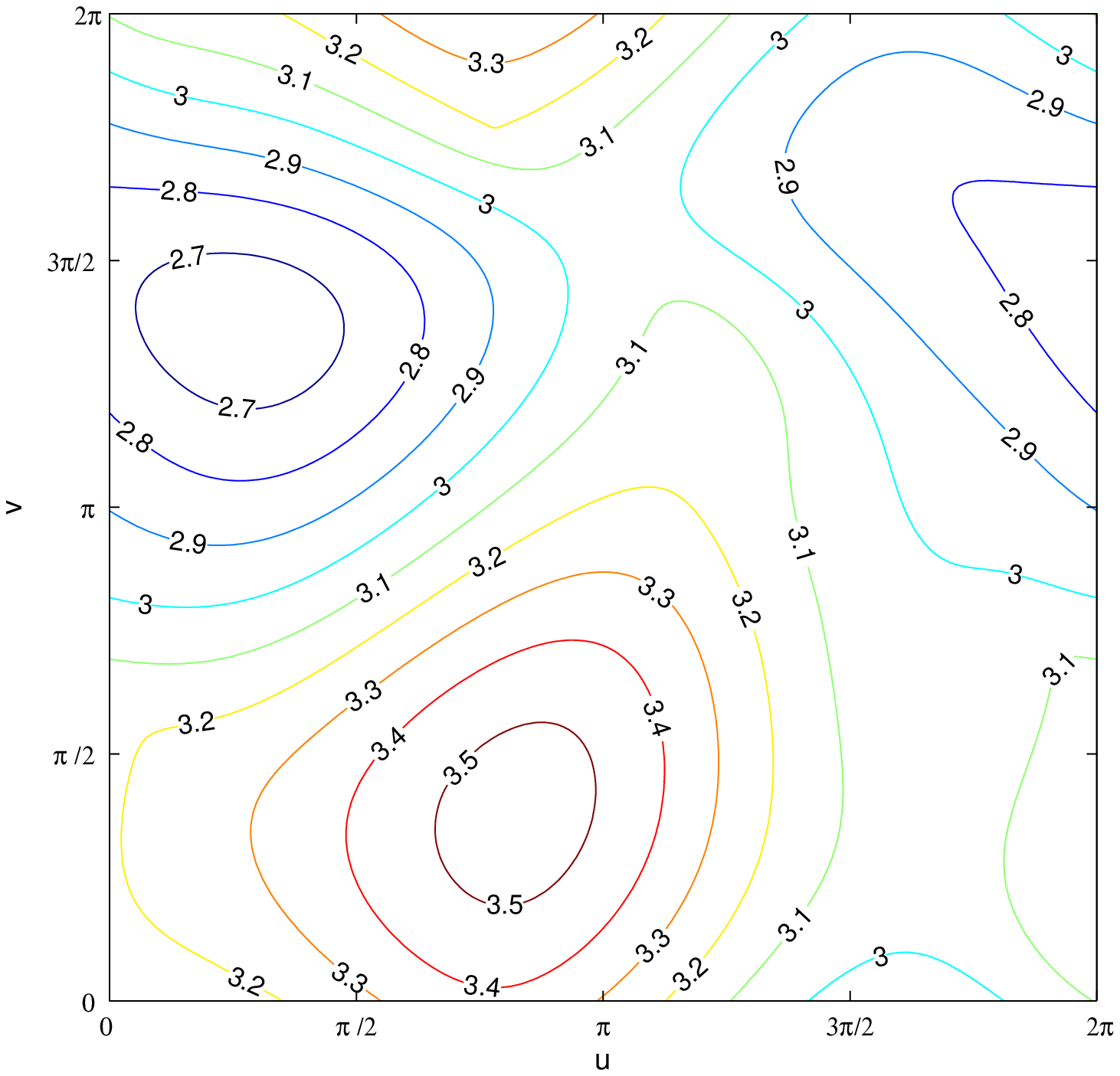} 
\includegraphics[viewport=260 25 700 470,width=8.5cm]{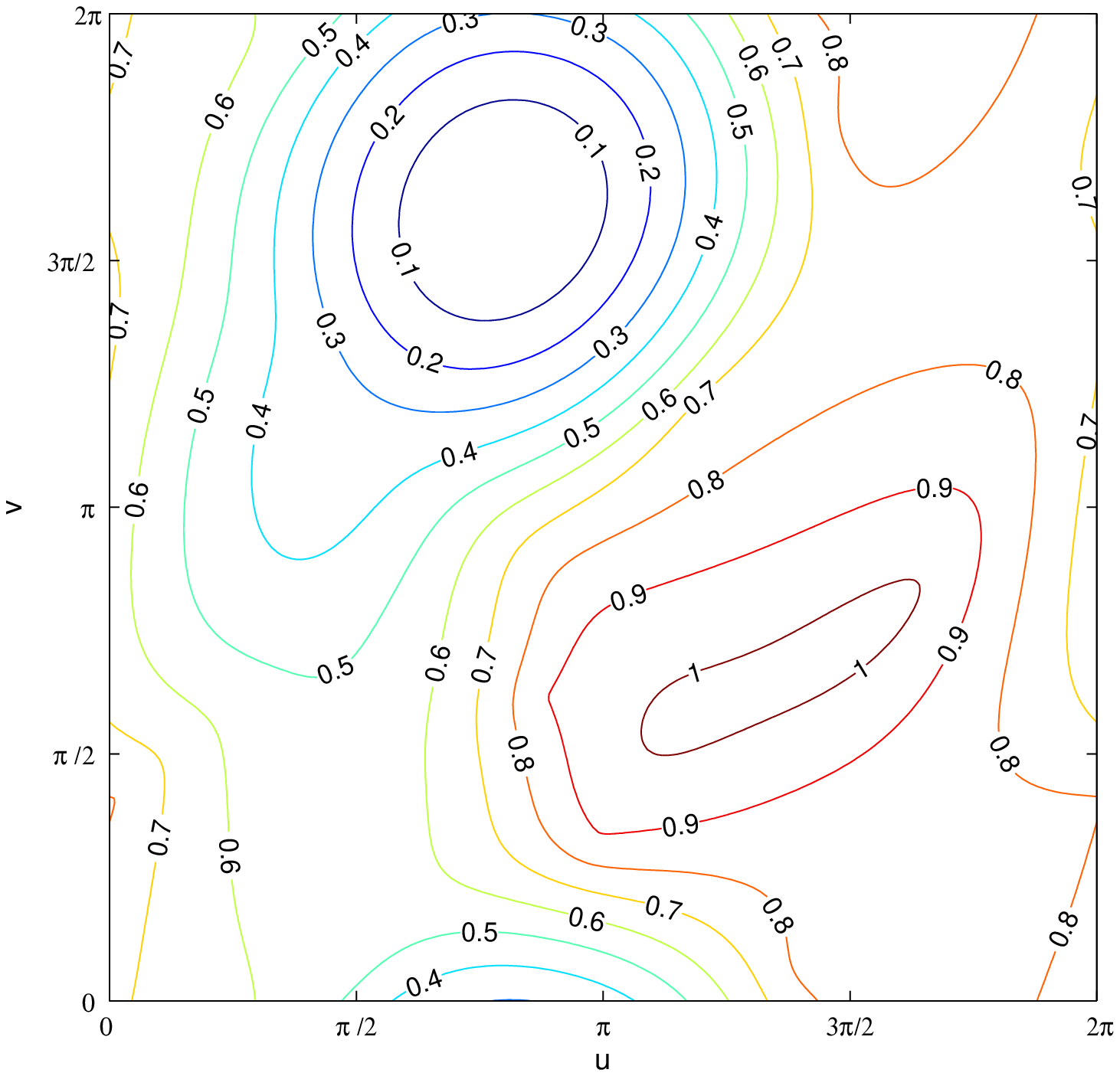} 
\end{center}
\caption{Contour plots of the function $F(\vec{\alpha})$ in the $uv$-plane (see text) of the constellation space ${\cal C}_6(5,4^2,2)$ in dimension six  (top), and of the constellation space ${\cal C}_5(4^3,2)$ in dimension five near a zero indicating a MU constellation $\{ 4^3,2\}_5$ (bottom).}
\label{slices}
\end{figure}

The function $F( \vec{\alpha} )$ achieves its maximum 
\be \label{Fmax}
F_{\mbox{\tiny max}} 
  = \frac{1}{2}\sum_{b=1}^d x_{b} (x_{b} - 1) + \left( \frac {\sqrt{d}-1}{\sqrt{d}} \right)^{2}  \sum_{1\leq b<b^\prime}^d x_{b} x_{b^\prime} \, ,
\ee
if all states coincide, each having components equal to $1 / \sqrt{d}$ only. For typical constellations such as $\{5^2,4,1\}_6$ or $\{5,3^3\}_6$, one finds $F_{\mbox{\tiny max}} = 33.2$ and $F_{\mbox{\tiny max}} = 25.0$, respectively. The top image of Fig. \ref{slices} shows a two-dimensional contour plot of $F(\vec{\alpha})$ in the 45-dimensional constellation space ${\cal C}_6 (5,4^2,2)$. Ranging between 2.6 and 3.6, the function $F(\vec{\alpha})$ exhibits one maximum, one minimum, and two saddle points. This structure is consistent with (\ref{function}) because $F(\vec{\alpha})$ reduces to a simple trigonometric polynomial of two variables if all but the first two angles $\alpha_1 \equiv u, \alpha_2\equiv v$, are fixed.  

Considering the range of the function $F$, it appears reasonable to say that a MU constellation $[x]_d$ parameterized by $\vec{\alpha}$ has been found if $F(\vec{\alpha})$ assumes a value below 
\be \label{definezero}
F_{\mbox{\tiny c}} = 10^{-7} \, .
\ee
This criterion, stronger than the one used in \cite{butterley+07} is entirely arbitrary, and smaller values could be used at the expense of computational time. The numerical data presented below will retrospectively justify the chosen value of the threshold for zeros of F.

\subsection{Testing the numerical search} \label{testsearches}

We begin by presenting searches for MU constellations which are known to exist. The data provide evidence that the numerical minimization of $F(\vec{\alpha})$ defined in Eq. (\ref{function}) is a reliable tool to identify MU constellations. 

\subsubsection{Three complete MU bases\label{threecompletebases}}

It is known that one can construct \emph{three} complete MU bases in the space $\mathbb{C}^d$ without refering to the prime decomposition of $d$ \cite{grassl04}. Let us check the proposed minimisation method by searching for the MU constellations $\{(d-1)^3\}_d$ in dimensions $d=2,3, \ldots,8$. Table \ref{threeMUbases100}  displays the success rates obtained for a total of 1,000 searches in each of these dimensions. The input consists of constellations $[(d-1)^3]_d$ chosen randomly in the constellation space ${\cal C}_d((d-1)^3) \equiv [0,2\pi)^{p_d}$, with $p_d = (d-1)(2d-3)$. Each dephased constellation $[(d-1)^3]_d$ corresponds to $3(d-1)$ pure states in $\mathbb{C}^d$.  

The searches are successful in all dimensions. The rate of success systematically decreases for larger dimensions if even and odd dimensions are considered separately. This overall trend is not surprising in view of the constant number of samples taken in ever bigger spaces ${\cal C}_d$. The success rate is consistently higher in even dimensions which might be attributed to the possibility of constructing different types of triples of MU bases resulting from the factor of two in $d=4,6,8$.   

\subsubsection{MU constellations in dimension five} \label{MUconst in d=5}

Next, we test the minimisation procedure by systematically searching for MU constellations of the form $\{4,x,y,z\}_5$, i.e. all MU constellations in dimension $d=5$ contained in four MU bases. The results from 1,000 searches for each MU constellation have been collected in Table \ref{dim5}. The success rate gradually decreases from 100\% for MU constellations with 16 or fewer parameters to 10\% for MU constellations with 44 parameters. All MU constellations are identified. In view of later developments the table also makes explicit the number of free parameters for each dephased constellation.

\begin{figure}[ht]
\begin{center}
\includegraphics[width=13cm]{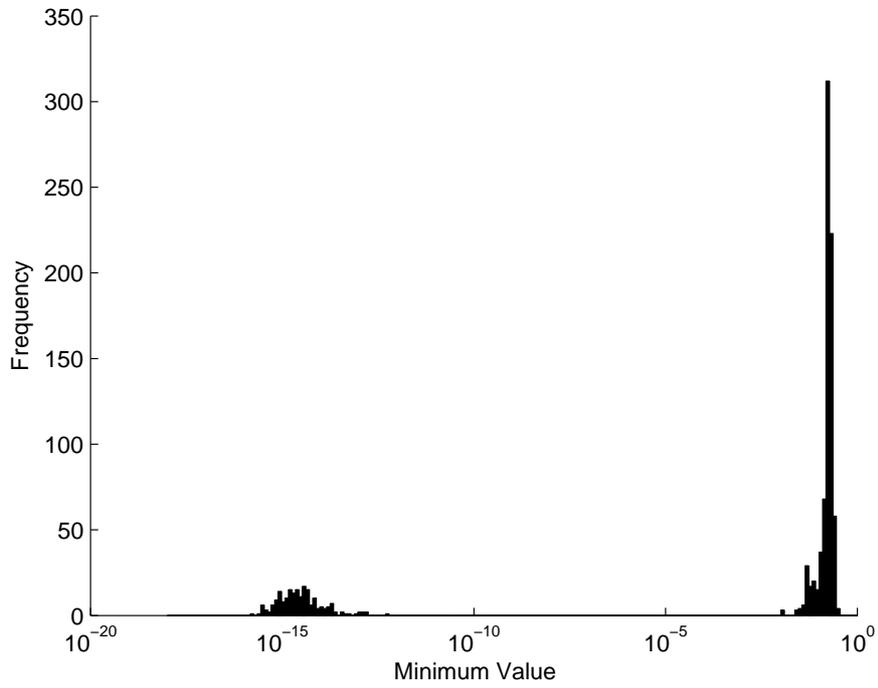}
\includegraphics[width=13cm]{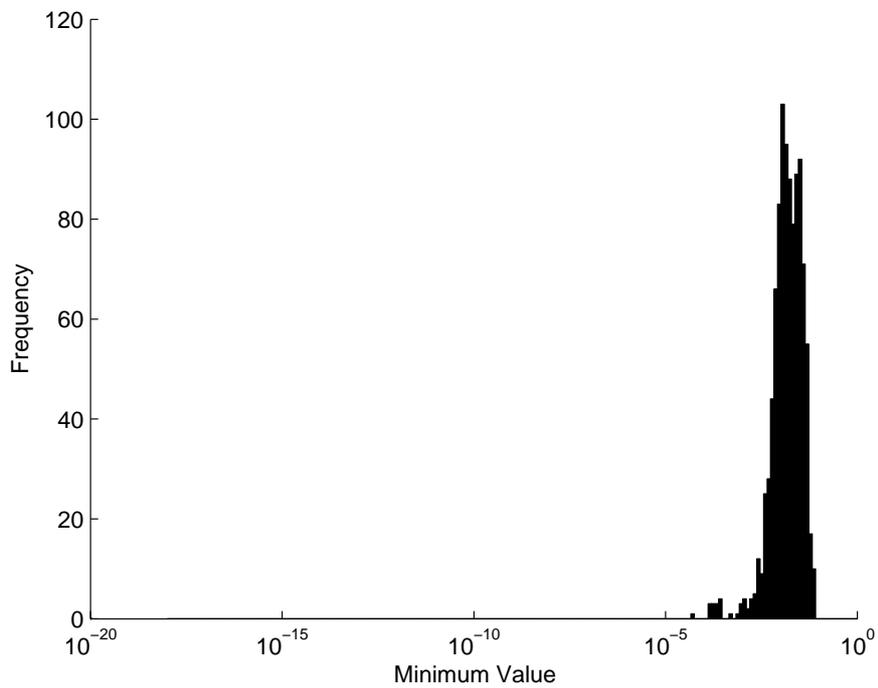} 
 
\end{center}
\caption{Distribution of the values obtained by minimising the function $F(\vec{\alpha})$ for 1,000 initial points chosen randomly in the 36-dimensional space $ {\cal C}_5(4^3,2)$ (top) and in the 45-dimensional constellation space ${\cal C}_6(5,4^2,2)$ (bottom).}
\label{hist442}
\end{figure}

To judge the quality of the minimisation procedure, it is instructive to plot the distribution of the minimal values of $F(\vec{\alpha})$ obtained in the space ${\cal C}_5( 4^3,2)$, say. The histogram at the top of Fig. \ref{hist442} shows that \emph{global} minima, defined by $F < 10^{-7}$, are separated from \emph{local} minima by several orders of magnitude, justifying the criterion (\ref{definezero}).
For a random sample of these 'zeros,' we have been able to reduce the value of $F(\vec{\alpha})$ to less than $10^{-20}$, simply by running the search for longer.  

Note that by detecting one MU constellation in a particular run, all MU constellations contained in it have also been found. Thus, Table \ref{dim5} does not only report 370 incidences of the MU constellation $\{ 4^2,2^2\}_5$ but since MU constellations form a lattice due to (\ref{partialorder}), all successful searches to the right and below this entry also confirm its presence, adding a further 983 detected cases.

The bottom image of Fig. \ref{slices} plots the contours of the function $F(\vec{\alpha})$ in a two-dimensional neighbourhood of a zero, i.e. of a MU constellation of type $\{ 4^3,2\}_5$. Qualitatively, it resembles the random cross-section depicted above it.   

\subsubsection{MU constellations in dimension seven} \label{MUconst in d=7}

In dimension seven, a complete set of eight MU bases exists. Thus, we expect a numerical search to successfully identify all MU constellations with no more than four partial bases. The largest constellation, $\{6^4\}_7$, now depends on 102 parameters, more than double the number occurring in dimension five. Due to this substantial expansion of the parameter space, however, the search for zeros of the function $F(\vec{\alpha})$ is likely to succeed less frequently. 

These expectations are confirmed by the results collected in Table \ref{dim7}. As in dimension five, the success rates decreases if MU constellations containing more states are being searched for. Although the spaces searched are considerably larger, we still find four out of five MU constellations of the form $\{6,x,y,z\}_7$ after 1,000 attempts. Overall, the success rates show a structure different from the one observed in dimension five: the high detection rate for small MU constellations drops sharply when the constellations approach any of the critical constellations $\{{\bf x}\}_7$. Importantly, all but one of the overdetermined MU constellations $\{\underline{x}\}_7$ beyond the `line' of critical MU constellations have been identified. It is true that the success rate is small but the basin of attraction for global minima is likely to be only a tiny region in the high-dimensional search space. 

\begin{figure}[ht]
\begin{center}
\includegraphics[width=13cm]{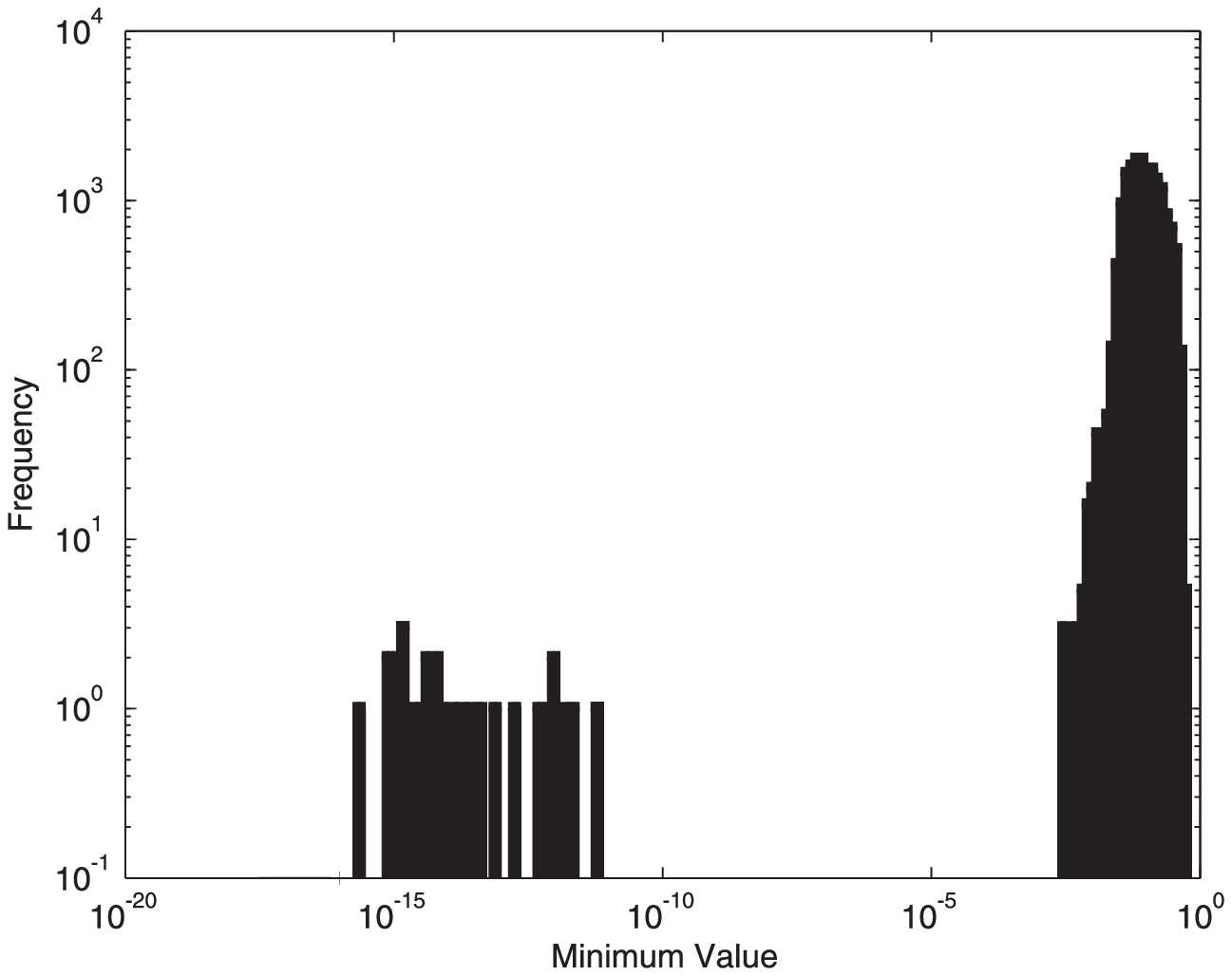}
\includegraphics[width=13cm]{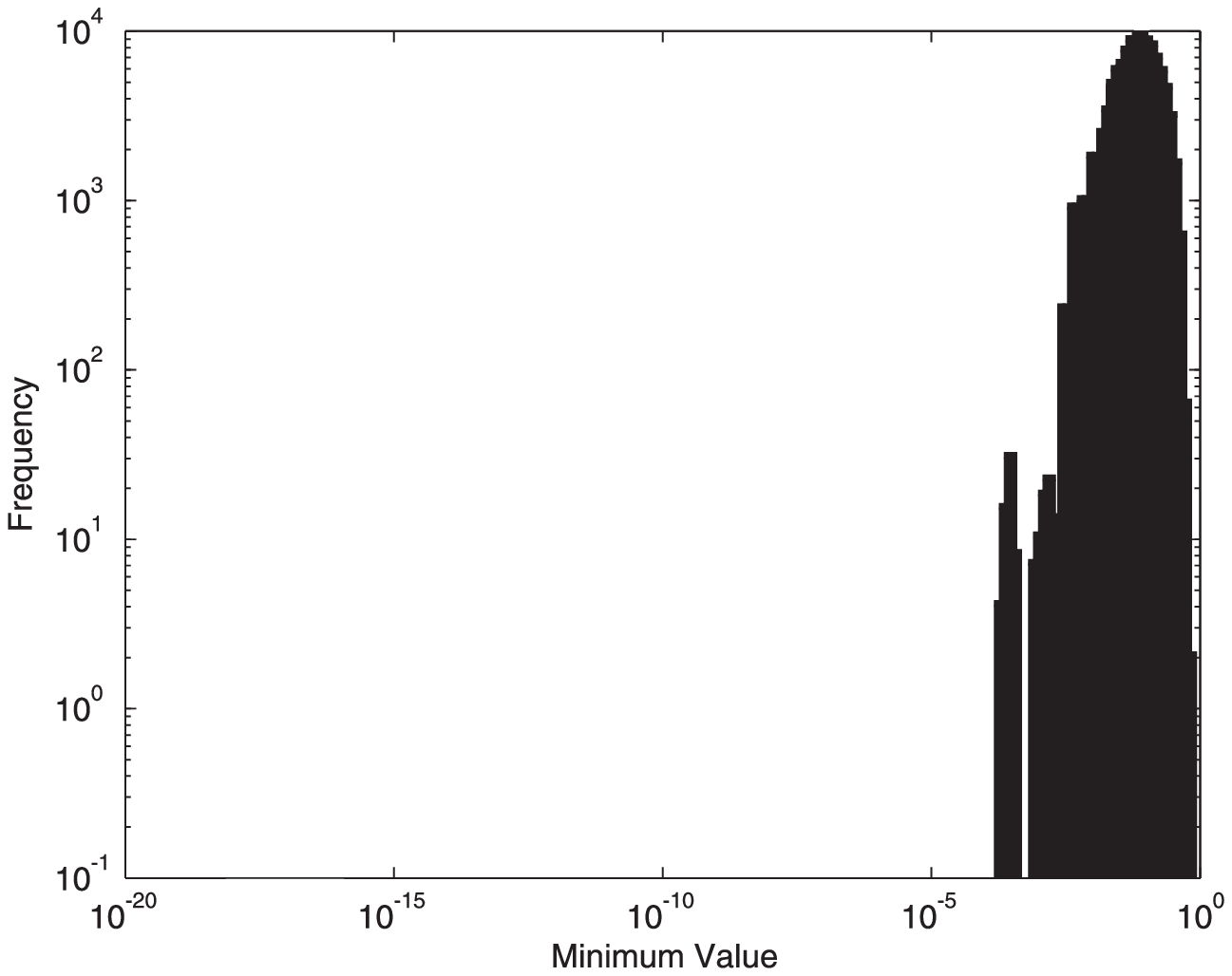} 
 
\end{center}
\caption{Distribution of the values obtained by minimising the function $F(\vec{\alpha})$ for 16,000 points combining the results of the 16 overdetermined constellations $ \{\underline{x}\}_7$ in Table \ref{dim7} (top), and for 110,000 points combining the results of the 11 overdetermined constellations $ \{\underline{x}\}_6$ in Table \ref{dim6} (bottom).}
\label{loghist}
\end{figure}

The quality of the zeros is excellent: they correspond to values of $F(\vec{\alpha})$ below $10^{-12}$, being clearly different from the vast majority of local minima producing values in the order of $10^{-3}$. This is illustrated in the upper histogram of Fig. \ref{loghist} which combines all the minima obtained for overdetermined constellations $\{ \underline{x} \}_7$. We associate the clusters of values at $10^{-13}$ and at $10^{-3}$ with global and local minima, respectively. 

It is straightforward to check that the numerically identified MU constellations reproduce the numbers $\chi _{jj^\prime}^{bb^\prime }$ in (\ref{definitionofchi}), correct to seven significant digits. We are thus confident to have identified these overdetermined MU constellations in dimension seven.    

\section{MU constellations in dimension six \label{Results}}

Knowing that the numerical procedure to minimise $F(\vec{\alpha})$ defined in (\ref{function}) generates reliable data, we now turn to the main findings of this paper which are related to dimension six.  

In Table \ref{dim6}, we present the success rates to identify all MU constellations contained in four MU bases $\{ 5^4\}_6$, i.e.
\be
\{5,x,y,z\}_{6}\, , \quad 1 \leq x,y,z\leq d-1 \, .
\ee
We will proceed as in dimensions $d=5$ and $d=7$ but, in order to give our results additional weight, we have performed 10,000 searches for each MU constellation.

The results exhibit a structure which differs qualitatively from the findings in neighboring dimensions. The success rates decrease as before if the search aims at MU constellations with increasing numbers of free parameters. However, after dropping to zero on the line of critical constellations $\{{\bf x}\}_6$, there is no evidence  for a single overdetermined MU constellation $\{\underline{x}\}_6$. 

It is true that only a few of these MU constellations had been identified in dimension seven; considering their abundance in $d=5$, however, their complete absence in $d=6$ is a striking feature which we consider to be statistically relevant. Note that the lattice structure due to (\ref{partialorder}) allows us to conclude that unsuccessful searches for MU constellations \emph{contained} in $\{ 5^4\}_6$ also count against its existence. Since none of the constellations it contains have been found, Table \ref{dim6} effectively reports a total of 170,000 \emph{negative} instances for the MU constellation $\{ 5^4 \}_6$.   

The minimal values of $F(\vec{\alpha})$ obtained for most of the constellations on and near the critical line are not below $1.1 \times 10^{-4}$ except for $\{ 5,4,3,2 \}_6, \{5,4^2,2\}_6$, and $\{5,3^3\}_6$, where values close to $10^{-6}$ have been obtained. We have not been able to push the these values below the threshold of $10^{-7}$, even by running the search considerably longer. The bottom histogram Fig. \ref{hist442} shows that the minima obtained for $ \{ 5,4^2,2 \}_6$ cluster at values of $10^{-3}$, orders of magnitude away from the criterion (\ref{definezero}) for a global minimum. The second histogram in Fig. \ref{loghist} combines the results for all overdetermined constellations $\{ \underline{x} \}_6$, showing that throughout the minimal values found are well above  the threshold of $10^{-7}$.

As an aside, the absence of the MU constellation $\{5^{2},4,1 \}_6$ from Table \ref{dim6} suggests, by the inclusion $ \{5^{3},1\}_6$ $\geq$ $ \{5^{2},4,1 \}_6$, that no three complete MU bases plus one additional mutually unbiased state exist. This result generalizes the impossibility to extend two MU bases $\{ 5^2 \}_6$ related to the Heisenberg-Weyl group to a MU constellation $\{5^{3},1\}_6$ \cite{grassl04}.

\section{Summary and Discussion}

We have defined constellations of quantum states in the space $\mathbb{C}^d$ which are mutually unbiased.  The search for these MU constellations has been cast in the form of a global minimisation problem which can be approached by standard numerical methods. Our conclusions are based on a total of 433,000 searches in dimensions five to seven which would take approximately 16,000 hours on a single Pentium 4 desktop PC. 

The results of the searches performed in dimension six provide strong evidence that not all MU constellations of the form $\{ 5,x,y,z\}_6 $ exist. Here are our main conclusions drawn from Table \ref{dim6}: 

\begin{itemize}
\item the \emph{largest existing} MU constellations are $\{ 5,4^2,1\}_6$ and $\{ 5^2,3,1\}_6$ both containing 15 ($\equiv S+1$) mutually unbiased states; 
\item the \emph{smallest non-existing} MU constellations are $\{5,3^3\}_6$ and $\{ 5,4,3,2\}_6$ each consisting of 14 ($\equiv S$) states; 
\item only \emph{one critical} MU constellation $\{ {\bf x}\}_6$ exists, namely $\{ 5^3\}_6$ corresponding to three MU bases with 18 states;
\item \emph{no overdetermined} MU constellation $\{ \underline{x}\}_6$ exists.
\end{itemize}
We have been able to positively identify 18 out of 35 MU constellations in dimension six. On the basis of the numerical data, we consider it highly unlikely that the 15 unobserved critical and overdetermined MU constellations do exist, making the existence of four MU bases exceedingly improbable. 

Let us discuss these results in a general framework. \emph{Critical} constellations $[ {\bf x}]_d$ have been defined by the equality $p_d = c_d$. If $p_d$ parameters need to satisfy $c_d \equiv p_d$ equations, one would expect some isolated solutions to exist in a generic situation. As three MU bases are critical constellations in any dimension $d$, they are expected to exist generically. In the overdetermined case, there are more constraints than free parameters, $c_d > p_d$, and no MU constellations are expected. This observation agrees with the fact that one can actually construct three MU bases without referring to the decomposition of $d$ into its prime factors.

The counting of parameters indicates how special large sets of mutually unbiased states are. For any $d>2$, the $d(d+1)$ quantum states of a complete set of MU bases possess too few parameters to generically satisfy the conditions imposed on them by mutual unbiasedness. In dimension seven, for example, such a set consists of 56 pure states depending on 288 independent parameters which need to satisfy 1176 constraints. This is only possible if the constraints conform to some fundamental structure prevailing in the space $\mathbb{C}^7$---obviously, the  number-theoretic consequences of $d=7$ being a prime number spring to mind. In other words, the constraints must \emph{degenerate} at one or more points of the constellation space ${\cal C}_7$ so that sufficiently many MU bases can arise. This observation could explain why Table \ref{dim7} seems to say that overdetermined constellations are easier to detect than critical ones: beyond the critical constellations $[ {\bf x} ]_7$, the additional free parameters in a constellation $[ \underline{x} ]_7$ might make it easier to locate the points of ${\cal C}_7$ where the constraints degenerate.

The implications of counting parameters apply not only to $d=2$ where precisely three MU bases exist but they also agree with the data in Table \ref{dim6}: not a single overdetermined MU constellation of the form $\{5,x,y,z\}_6$ has been observed. Thus, it is natural to suspect that \emph{all} overdetermined MU constellations $\{\underline{x}\}$ will be missing in dimension six. More generally, suppose it is the smallest prime in the decomposition of $d$ that limits the number of MU bases. Then, for dimensions that contain only a single factor of two, we also expect that no overdetermined MU constellations $\{ \underline{x}\}_{2d}$ exist. For example, we consider it unlikely in dimension ten to find MU constellations of the form $\{ 9,x,y,z \}_{10}$ with $x+y+z =18$.   

We conclude by emphasizing that the results of the numerical searches presented in Table \ref{dim6} provide strong evidence for the absence of seven MU bases in dimension six. It is thus likely that the kinematics of quantum systems with dimensions $d_1 = 2 \times 3$ and $d_2 = 3 \times 3$, respectively, will differ structurally. Due to the relations established in \cite{boykin+05}, the data collected also support the conjecture \cite{kostrikin+94} that the Lie algebra $sl_6(\mathbb{C})$ has no orthogonal decomposition.

\subsection*{Acknowledgments}

Discussions with the participants of the Quantum Information Seminar at the Department of Mathematics at the University of York are gratefully acknowledged, and with Tony Sudbery in particular. We would also like to thank Subhash Chaturvedi for asking if seven mutually unbiased states exist in dimension six (they do, as does the MU constellation $\{2^7\}_6$).

\newpage

\appendix

\section{Equivalence classes of MU constellations}

Two complete sets of MU bases $\{ {\cal B}_0, \ldots , {\cal B}_d \}$ and $\{ {\cal B}_0^\prime , \ldots , {\cal B}_d^\prime \}$ are said to be equivalent,
\be \label{MUequivalence}
\{ {\cal B}_0, \ldots , {\cal B}_d \} \cong \{ {\cal B}_0^\prime , \ldots , {\cal B}_d^\prime \},
\ee
if we can obtain one from the other by a succession of the following four transformations:

\begin{enumerate} 
\item an \emph{overall unitary} transformation $U$, 
\be \label{overallunitary}
U \{{\cal B}_{0},\ldots ,{\cal B}_{d}\} 
 \equiv \{U{\cal B}_{0},\ldots ,U{\cal B}_{d}\} 
 \cong \{{\cal B}_{0},\ldots ,{\cal B}_{d}\} \, , 
\ee
which leaves invariant the value of all scalar products;
  
\item $(d+1)$ simultaneous unitary transformations $D_b$ which multiply each vector with a phase factor,
\be \label{simdiagunitaries}
\{ {\cal B}_{0}D_{0},\ldots ,{\cal B}_{d} D_{d}\} \cong \{{\cal B}_{0},\ldots ,{\cal B}_{d}\} \, , 
\ee
exploiting the fact that the physically irrelevant overall phase of a quantum state drops out of the conditions (\ref{MUB conditions});

\item pairwise exchanges of any two bases,
\be \label{swaps}
\{ \ldots , {\cal B}_b, \ldots , {\cal B}_{b^\prime}, \ldots \} 
\cong 
\{ \ldots , {\cal B}_{b^\prime}, \ldots , {\cal B}_b, \ldots \} \, ,
\ee
which amounts to relabeling the bases;
\item $(d+1)$ \emph{simultaneous permutations} $P_b$ of the members within each basis, 
\be \label{simpermutations}
\{{\cal B}_{0}P_{0},\ldots ,{\cal B}_{d}P_{d}\} \cong \{ {\cal B}_{0},\ldots , {\cal B}_{d} \}\, , 
\ee
which amounts to relabeling the elements of each basis.  
\end{enumerate}
For simplicity, we have written $U{\cal B}_b \equiv \{U \ket{\psi_1^b}, \ldots , U \ket{\psi_d^b}\}$, that is, the unitary $U$ acts on each member of the basis ${\cal B}_b$; the expressions ${\cal B}_b D_b$ etc. are defined similarly.   
 
These equivalence relations can be used to \emph{dephase} a given complete set of MU bases $\{ {\cal B}_0, \ldots , {\cal B}_d \}$. Written in dephased form, its first basis is given by the standard basis ${\cal B}_z$, the components of the first vector of the second basis are equal to $1/\sqrt{d}$, as are the first components of the remaining $(d-1)(d+1)$ vectors. Let us illustrate the dephasing in dimension $d=3$ where a given complete set of four MU bases can be brought into the form
\begin{eqnarray} \label{exampled=3}
\left\{ \left( 
\begin{array}{ccc}
1 & 0 & 0 \\ 
0 & 1 & 0 \\ 
0 & 0 & 1%
\end{array}%
\right) \, , \quad \frac{1}{\sqrt{3}}\left( 
\begin{array}{ccc}
1 & 1 & 1 \\ 
1 & e^{i\alpha_{11}} & e^{i\alpha_{12}} \nonumber \\ 
1 & e^{i\alpha_{21}} & e^{i\alpha_{22}}%
\end{array}%
\right) \, , \hspace{3cm} \right.\\ 
\left. \frac{1}{\sqrt{3}} 
\left( 
\begin{array}{ccc}
1 & 1 & 1 \\ 
e^{i\beta_{11}} & e^{i\beta_{12}} & e^{i\beta_{13}} \\ 
e^{i\beta_{21}} & e^{i\beta_{22}} & e^{i\beta_{23}}%
\end{array}%
\right) \, ,  \quad
\frac{1}{\sqrt{3}}\left( 
\begin{array}{ccc}
1 & 1 & 1 \\ 
e^{i\gamma_{11}} & e^{i\gamma_{12}} & e^{i\gamma_{13}} \\ 
e^{i\gamma_{21}} & e^{i\gamma_{22}} & e^{i\gamma_{23}}%
\end{array}%
\right) \right\} \, .
\end{eqnarray}
The three orthonormal states of each basis have been arranged into four unitary matrices. The second unitary matrix obtained here is (proportional to) a  \emph{dephased} complex Hadamard matrix \cite{bengtsson+07} motivating our terminology. Note that the vectors of the last three bases (except for $(1,1,1)^T/\sqrt{3}$) may be rearranged using (\ref{simpermutations}).

To dephase a given set of $(d+1)$ MU bases, we first apply the overall unitary operator $U_1$ chosen in such a way that ${\cal B}_{0}$ turns into the standard basis ${\cal B}_z$. This transformation simply corresponds to a change of basis in $\mathbb{C}^d$. As an immediate consequence of (\ref{MUB conditions}), the components of each vector of the remaining bases need to have modulus $1/\sqrt{d}$. Assuming that the first vector of the basis ${\cal B}_{1}$ is given by $(e^{i\delta_{1}},\ldots ,e^{i\delta_{d}})^T/\sqrt{d}$, we apply a second overall unitary operator $U_2 = \mbox{diag}(e^{-i\delta_{1}},\ldots
,e^{-i\delta_{d}})$ so that each component of the first vector of ${\cal B}_{1}$ now equals $1/\sqrt{d}$. Note that this transformation introduces additional phase factors on all other states including those of ${\cal B}_0$. Finally, we use a transformation of type (\ref{simdiagunitaries}) with operators $D_b$ determined in such a way that ${\cal B}_0$ becomes the standard basis again, and that the first component of each state of ${\cal B}_1, \ldots , {\cal B}_{d}$ equals $1/\sqrt{d}$.

MU constellations $\{ x \}_d$ with at least one complete basis as in (\ref{restriction}) also come in equivalence classes if one applies suitably restricted variants of the symmetry transformations (\ref{overallunitary}) to (\ref{swaps}). Thus, they can be brought to \emph{dephased} form as well. If a complete set of MU bases exists, such as $\{ 3^4 \}_4$ in $\mathbb{C}^d$, the dephased form of smaller MU constellations is simply obtained by removing an appropriate number of the vectors. Eq. (\ref{exampled=3,2221}) shows the dephased form of the MU constellation $\{ 2^3,1\}_4$ contained in $\{ 3^4\}_4$, given in (\ref{exampled=3}). 

\clearpage

\section*{Figures}

\vspace{1.5cm}

\clearpage

\section*{Tables}

\vspace{2cm}

\begin{table}[ht]
\begin{center}
\begin{tabular}{|c||ccccccc|}
\hline 
               &  \multicolumn{7}{|c|}{\mbox{success rate}}\\ 
\hline
Dimension       & 2 & 3 & 4 & 5 & 6 & 7 & 8 \\ 
\hline 
$p_d$           & 1  & 6 & 15 & 28 & 45 & 66 & 91 \\
\hline \hline
$\{(d-1)^3\}_d$ & 100.0  & 81.9 & 96.6 & 49.3 & 67.9 & 24.0 & 48.5\\ 
\hline
\end{tabular}
\end{center}
\caption{Success rates for searches of three MU bases in dimensions $d=2, 3, \ldots , 8$, based on 1,000 initial points randomly chosen in the $p_d$-dimensional space ${\cal C}_d((d-1)^3)$.}
\label{threeMUbases100}
\end{table}

\vspace{3cm}

\begin{table}[hb]
\begin{center}
\begin{tabular}{|c|cccc||cccc|}
\hline 
 $d=5$       & \multicolumn{4}{|c||}{ parameters $p_5$} & \multicolumn{4}{|c|}{success rate}\\
 \hline 
$ x, y$ & \multicolumn{4}{|c||}{$z$} & \multicolumn{4}{|c|}{$z$}\\ 
		& 1 & 2 & 3 & 4 & 1 & 2 & 3 & 4 \\ 
\hline  \hline
1,1 &8&-&-&-& 100.0  &  -   & -  & -   \\ 
\hline
2,1 &12&-&-&-& 100.0  &  -   & -  & -   \\
2,2 &16&20&-&-& 100.0   &  96.4  & -  & -   \\ 
\hline
3,1 &16&-&-&-& 100.0   &  -   & -  & -   \\ 
3,2 &20&24&-&-& 92.0   &  35.7  & -  & -   \\
3,3 &24&\textbf{28}&\underline{32}&-& 68.3   &  \textbf{38.0}  & \underline{29.0} & -   \\  
\hline
4,1 &20&-&-&-& 99.0  &  -   & -  & -    \\  
4,2 &24&\textbf{28}&-&-& 56.2   &  \textbf{37.0}  & -  & -    \\
4,3 &\textbf{28}&\underline{32}&\underline{36}&-& \textbf{55.8}   &  \underline{31.8}  & \underline{21.8} & -    \\ 
4,4 &\underline{32}&\underline{36}&\underline{40}&\underline{44}& \underline{37.4}   &  \underline{20.1}  & \underline{14.9} & \underline{9.7}   \\ 
\hline
\end{tabular}
\end{center}
\caption{Success rates for searches of MU constellations $\{4,x,y,z\}_5$ in dimension five, based on 1,000 initial points randomly chosen in the $p_5$-dimensional space ${\cal C}_5(4,x,y,z)$; bold entries signal critical constellations $\{ {\bf x}\}_5$, while underlined entries correspond to overdetermined constellations $\{ \underline{x}\}_5$.}
\label{dim5}
\end{table}

\begin{table}[ht]
\begin{center}
\begin{tabular}{|c|cccccc||rrrrrr|}
\hline 
 $d=7$ & \multicolumn{6}{|c||}{\mbox{parameters $p_7$}} & \multicolumn{6}{|c|}{\mbox{success rate}}\\ 
 \hline
 $ x, y$ & \multicolumn{6}{|c||}{$z$} & \multicolumn{6}{|c|}{$z$}\\
         & 1    & 2 & 3 & 4 & 5 & 6 & 1   & 2 & 3 & 4 & 5 & 6\\ 
\hline \hline
1,1 &12 &  - &  - &  - &  - & - & 100.0 & -   & - & - & - & -\\ 
\hline
2,1 & 18 &  - &  - &  - &  - & - & 100.0& -   & - & - & - & -\\
2,2 & 24 & 30 &  - &  - &  - & - & 100.0 & 100.0 & - & - & -& -\\ 
\hline
3,1 &24  &  - &  - &  - &  - & - & 100.0 & -  & - & - & - & -\\ 
3,2 &30 &36 &  - &  - &  - & - & 100.0 & 100.0& - & - & - & - \\ 
3,3 &36 &42&48 &  - &  - & - & 100.0  & 100.0 & 99.3 & - & - & -\\ 
\hline
4,1 &30 &  - &  - &  - &  - & - &100.0 & -  & - & - & - & -\\ 
4,2 & 36 & 42 &  - &  - &  - & -& 100.0  & 100.0 & - & - & - & -\\
4,3 &42 &48 & 54 &  - &  - & - & 99.9   & 95.6  & 0.0 & - & - & -\\ 
4,4 &48 & 54 & 60 & \bf{66} &  - & - &  52.3   & 0.0  & 0.0 & \bf{0.0} & - & -\\ 
\hline 
5,1 &36 & -  & -  & -  &  - & - & 100.0 & -  & - & - & - & -\\ 
5,2 & 42 & 48 & -  & -  &  - & - & 100.0  & 37.9 & - & - & - & -\\ 
 
5,3 &48 & 54 & 60 & -  &  - & - &2.6   & 0.0  & 0.1 & - & - & -\\ 
5,4 & 54 & 60 & \bf{66} & \underline{72} &  - & - & 0.0   & 0.0  & \bf{0.0} & \underline{0.1} & - & -\\
5,5 & 60 & \bf{66} & \underline{72} & \underline{78} & \underline{84} & - & 0.2   & \bf{0.2}  & \underline{0.2} & \underline{0.1} & \underline{0.2} & -\\ 
\hline  
6,1 &42 & -  & -  & -  &  - & - & 57.5 & -  & - & - & - & -\\ 
6,2 & 48 & 54 & -  & -  &  - & - & 1.1 & 0.0 & - & - & - & -\\ 
 
6,3 &54 & 60 & \bf{66} & -  &  - & - &  0.0 & 0.1  & \bf{0.0} & - & - & -\\ 
6,4 & 60 & \bf{66} & \underline{72} & \underline{78} &  - & - & 0.2   & \bf{0.0}  & \underline{0.1} & \underline{0.3} & - & -\\
6,5 & \bf{66} & \underline{72} & \underline{78} & \underline{84} & \underline{90} & - & \bf{0.3}   & \underline{0.4}  & \underline{0.1} & \underline{0.1} & \underline{0.1} & -\\
6,6 & \underline{72} & \underline{78} & \underline{84} & \underline{90} & \underline{96} & \underline{102} & \underline{0.5}   & \underline{0.2}  & \underline{0.2} & \underline{0.0} & \underline{0.4} & \underline{0.3}\\ 
\hline    
\end{tabular}
\end{center}
\caption{Success rates for searches of MU constellations $\{6,x,y,z\}_7$ in dimension seven, based on 1,000 initial points randomly chosen in the $p_7$-dimensional space ${\cal C}_7(6,x,y,z)$; bold entries signal critical constellations $\{ {\bf x}\}_7$, while underlined entries correspond to overdetermined constellations $\{ \underline{x}\}_7$.}
\label{dim7}
\end{table}

\begin{table}[ht]
\begin{center}
\begin{tabular}{|c|ccccc||rrrrr|}
\hline 
 $d=6$ & \multicolumn{5}{|c||}{\mbox{parameters $p_6$}} & \multicolumn{5}{|c|}{\mbox{success rate}}\\ 
 \hline
 $ x, y$ & \multicolumn{5}{|c||}{$z$} & \multicolumn{5}{|c|}{$z$}\\
         & 1    & 2 & 3 & 4 & 5 & 1   & 2 & 3 & 4 & 5 \\ 
\hline \hline
1,1 &10 &  - &  - &  - &  - & 100.00 & -   & - & - & - \\ 
\hline
2,1 & 15 &  - &  - &  - &  - & 100.00& -   & - & - & - \\
2,2 & 20 & 25 &  - &  - &  - & 100.00 & 
100.00 & - & - & - \\ 
\hline
3,1 & 20 &  - &  - &  - &  - &100.00 & -  & - & - & - \\ 
3,2 &25 &30 &  - &  - &  - & 99.95 & 100.00 & - & - & - \\ 
3,3 &30 &35&40 &  - &  - &99.42  &39.03 & 0.00 & - & - \\ 
\hline
4,1 &25 &  - &  - &  - &  - &100.00 & -  & - & - & - \\ 
4,2 & 30 & 35 &  - &  - &  - & 92.92  & 44.84 & - & - & - \\
4,3 &35 &40 & \bf{45} &  - &  - & 12.97   & 0.00  & \bf{0.00} & - & - \\ 
4,4 &40 & \bf{45} & \underline{50} & \underline{55} &  - & 0.74   & \bf{0.00}  & \underline{0.00} & \underline{0.00} & - \\ 
\hline 
5,1 &30 & -  & -  & -  &  - & 95.40 & -  & - & - & - \\ 
5,2 & 35 & 40 & -  & -  &  - & 76.71  & 10.96 & - & - & - \\ 
 
5,3 &40 & \bf{45} & \underline{50} & -  &  - & 1.47   & {\bf 0.00}  & \underline{0.00} & - & - \\ 
5,4 & \bf{45} & \underline{50} & \underline{55} & \underline{60} &  - & \bf{0.00}   & \underline{0.00}  & \underline{0.00} & \underline{0.00} & - \\
5,5 & \underline{50} & \underline{55} & \underline{60} & \underline{65} & \underline{70} & \underline{0.00}   & \underline{0.00}  & \underline{0.00} & \underline{0.00} & \underline{0.00} \\ 
\hline   
\end{tabular}
\end{center}
\caption{Success rates for searches of MU constellations $\{5,x,y,z\}_6$ in dimension six, based on 10,000 initial points randomly chosen in the $p_6$-dimensional space ${\cal C}_6(5,x,y,z)$; bold entries signal critical constellations $\{ {\bf x}\}_6$, while underlined entries correspond to overdetermined constellations $\{ \underline{x}\}_6$.}
\label{dim6}
\end{table}


\begin{thebibliography}{99}

\bibitem{schwinger60}
J. Schwinger,
Proc. Nat. Acad. Sci. U.S.A., {\bf 46}, 560, (1960)

\bibitem{ivanovic81} I. D. Ivanovi\'c, J. Phys. A, {\bf 14}, 3241, (1981)

\bibitem{wootters+89}
W. K. Wootters and B. D. Fields,
Ann. Phys. (N.Y.), {\bf 191}, 363 (1989)

\bibitem{planat+06}
M. Planat, H. Rosu, S. Perrine, and M. Saniga,
Found. Phys., {\bf 36}, 1662 (2006)


\bibitem{butterley+07} P. Butterley and W. Hall,
Phys. Lett. A {\bf 369}, 5 (2007)

\bibitem{grassl04} M. Grassl, \emph{On SIC-POVMs and MUBs in Dimension 6}, in: \emph{Proc. ERATO Conference on Quantum Information Science (EQUIS 2004)}, J. Gruska (ed.)

\bibitem{nesterov00} Y. Nesterov: \emph{Squared functional systems and optimization problems}, in: \emph{High Performance Optimization}, H. Frenk, K. Roos, T. Terlaky, and S. Zhang (eds.), Kluwer, Dordrecht 2000




\bibitem{levenberg44} K. Levenberg, Quart. Appl. Math. {\bf 2}, 164 (1944)

\bibitem{marquardt63} D. W. Marquardt, J. Soc. Ind. App. Math. {\bf 11}, 431 (1963)

\bibitem{SciPy} E. Jones, T. Oliphant, P. Peterson et al. \emph{SciPy: Open Source Scientific Tools for Python 4.0} (Version 2.5), {\tt http://www.scipy.org/}


\bibitem{boykin+05} P.O. Boykin, M. Sitharam, P.H Tiep and P. Wocjan, Quantum Inf. Comp. {\bf 7}, 371 (2007)

\bibitem{kostrikin+94} A.I. Kostrikin and P.H. Tiep: \emph{Orthogonal Decompositions and Integral Lattices}, Walter de Gruyter, 1994.  

\bibitem{bengtsson+07} I. Bengtsson, W. Bruzda, \r{A}. Ericsson, J-\r{A}. Larsson, W.
Tadej and K. \.Zyczkowski, J. Math. Phys. {\bf 48}, 052106 (2007)
\end{thebibliography}
\end{document}